\documentclass[11pt]{article}

\usepackage{amsmath}
\usepackage{graphicx}
\usepackage{amsfonts}
\usepackage{amssymb}
\usepackage{epsfig}
\usepackage{color}
\usepackage{psfrag}
\usepackage{epstopdf}

\newcommand{\EQ}{\begin{equation}}
\newcommand{\EN}{\end{equation}}
\newcommand{\be}{\begin{equation}}
\newcommand{\ee}{\end{equation}}
\newcommand{\bea}{\begin{eqnarray}}
\newcommand{\eea}{\end{eqnarray}}

\begin{document} \setcounter{page}{0}
\topmargin 0pt
\oddsidemargin 5mm
\renewcommand{\thefootnote}{\arabic{footnote}}
\newpage
\setcounter{page}{0}
\topmargin 0pt
\oddsidemargin 5mm
\renewcommand{\thefootnote}{\arabic{footnote}}
\newpage
\begin{titlepage}
\begin{flushright}
SISSA 60/2010/EP \\
\end{flushright}
\vspace{0.5cm}
\begin{center}
{\large {\bf On three-point connectivity in two-dimensional percolation}}\\
\vspace{1.8cm}
{\large Gesualdo Delfino$^{a,b}$ and Jacopo Viti$^{a,b}$}\\
\vspace{0.5cm}
{\em ${}^a\,$International School for Advanced Studies (SISSA), via Bonomea 265, 34136 Trieste, Italy}\\
{\em ${}^b\,$Istituto Nazionale di Fisica Nucleare, sezione di Trieste, Italy}\\
\end{center}
\vspace{1.2cm}

\renewcommand{\thefootnote}{\arabic{footnote}}
\setcounter{footnote}{0}

\begin{abstract}
\noindent 
We argue the exact universal result for the three-point connectivity of critical percolation in two dimensions. Predictions for Potts clusters and for the scaling limit below $p_c$ are also given.  
\end{abstract}
\end{titlepage}

\newpage
{\bf 1.} The connectivity functions $P_n(x_1,\ldots,x_n)$, i.e. the probabilities that $n$ points belong to the same finite cluster, play a fundamental role in percolation theory \cite{SA,Grimmett}. In particular, their scaling limit determines the universal properties that clusters exhibit near the percolation threshold $p_c$. Although most quantitative studies focus on the two-point connectivity, which determines observables as the mean cluster size $S=\sum_xP_2(x,0)$, also the connectivities with $n>2$ carry essential information about the structure of the theory. In this paper we consider the case $n=3$ for clusters in random percolation and in the $q$-state Potts model ($q\leq 4$) in two dimensions, in the scaling limit on the infinite plane. In particular, we will determine exactly the universal quantity
\EQ
R(x_1,x_2,x_3)=\frac{P_3(x_1,x_2,x_3)}{\sqrt{P_2(x_1,x_2)P_2(x_1,x_3)P_2(x_2,x_3)}}
\label{r}
\EN
at $p_c$, as well as some of its features below $p_c$. The connectivity combination (\ref{r}) was studied in \cite{KSZ,SKZ}  for the case in which two of the three points are located on the boundary of the half-plane, it was shown to be a constant at $p_c$ and determined numerically and also analytically, exploiting the fact that points on the boundary yield linear differential equations for the connectivities at criticality \cite{Cardy-crossing}. It is not known how to write differential equations for random percolation connectivities if all the points are in bulk.

It is well known that random percolation can be described as the limit $q\to 1$ of the $q$-state Potts model with lattice hamiltonian \cite{Potts,Wu}
\EQ
{\cal H}=-J\sum_{<x,y>}\delta_{s(x),s(y)}\,,\hspace{1cm}s(x)=1,\ldots,q,
\label{hamiltonian}
\EN
as a consequence of the fact that the partition function $Z=\sum_{\{s(x)\}}e^{-\cal H}$ can be rewritten as \cite{KF} 
\EQ
Z=\sum_G p^{n_b}(1-p)^{\bar{n}_b}q^{N_c}\,,
\label{rcm}
\EN
where the sum is taken over all graphs $G$ obtained putting bonds on the lattice, $n_b$ is the number of bonds in the graph, $\bar{n}_b$ the number of absent bonds, $p=1-e^{-J}$ the probability that a bond is present, and $N_c$ the number of connected components (clusters) in $G$, with isolated sites also counted as clusters. Notice that (\ref{rcm}) provides a continuation of the Potts model to real values of $q$. The random percolation transition corresponds to the limit $q\to 1$ of the ferromagnetic Potts transition, which is second order as long as $q$ does not exceed a value $q_c$ ($q_c=4$ in two dimensions  \cite{Baxter}). The connectivity functions can be related to the correlation functions of the Potts spin variable $\sigma_\alpha(x)=\delta_{s(x),\alpha}-1/q$, $\alpha=1,\ldots,q$; for $p\leq p_c$, namely when the probability of finding an infinite cluster is zero, the relation for the two- and three-point functions reads
\bea
&& P_2(x_1,x_2)=\frac{q^2}{q-1}\,\langle\sigma_\alpha(x_1)\sigma_\alpha(x_2)\rangle\,,
\label{p2}\\
&& P_3(x_1,x_2,x_3)=\frac{q^3}{(q-1)(q-2)}\,\langle\sigma_\alpha(x_1)\sigma_\alpha(x_2)\sigma_\alpha(x_3)\rangle\,,
\label{p3}
\eea
where for random percolation the limit $q\to 1$ is understood in the r.h.s. It can be shown \cite{CP}, however, that these relations, evaluated for a generic $q$, give the connectivities for the clusters entering the expansion (\ref{rcm}), the so called Kasteleyn-Fortuin (KF) clusters (or "droplets" \cite{SA}) of the Potts model.

It will be relevant for our purposes that in two dimensions the Potts model exhibits a duality property \cite{Potts} relating a value $K\equiv e^J-1$ of the coupling to the value $K^*=q/K$. The self-dual point $K_c=\sqrt{q}$ is the critical point. Duality allows to relate the correlation functions of the spin variables computed at a value $J$ of the coupling to the correlation functions of ``disorder'' variables $\mu_{\alpha\beta}$, $\alpha,\beta=1,\ldots,q$, $\alpha\neq\beta$, computed at the dual value $J^*$. In particular one obtains the relations
\bea
&& \langle\sigma_\alpha(x_1)\sigma_\alpha(x_2)\rangle_{J\leq J_c}=\frac{1}{q^2}(q-1)\,\langle\mu_{\alpha\beta}(x_1)\mu_{\beta\alpha}(x_2)\rangle_{J^*}\,,
\label{d2}\\
&& \langle\sigma_\alpha(x_1)\sigma_\alpha(x_2)\sigma_\alpha(x_3)\rangle_{J\leq J_c}=\frac{1}{q^3}(q-1)(q-2)\,\langle\mu_{\alpha\beta}(x_1)\mu_{\beta\gamma}(x_2)\mu_{\gamma\alpha}(x_3)\rangle_{J^*}\,.
\label{d3}
\eea
Comparison with (\ref{p2}-\ref{p3}) shows that $P_2$ and $P_3$ are related to the disorder correlators without any $q$-dependent prefactor. 

In the scaling limit ($q\leq q_c$, $J\to J_c$, distances $|x_i-x_j|\equiv r_{ij}$ much larger than the lattice spacing) that we consider from now on, the lattice variables become fields of the Potts field theory, i.e. the simplest field theory realizing the $S_q$ permutational symmetry characteristic of the hamiltonian (\ref{hamiltonian}). Up to overall constants depending on the arbitrary normalization of the field $\sigma_\alpha(x)$, the correlators in (\ref{p2}-\ref{p3}) are universal functions of the distances $r_{ij}$ measured in units of the connectivity length. The normalization constants cancel in the combination (\ref{r}), which is then completely universal.

\vspace{.3cm}
{\bf 2.} At criticality, conformal invariance implies that the two- and three-point correlators of scalar fields $A_i(x)$ with scaling dimension $X_i$ have the form \cite{Polyakov}
\bea
&& \langle A_1(x_1)A_2(x_2)\rangle=\delta_{X_1,X_2}C_{12}\,r_{12}^{-2X_1}\,,\\
&& \langle A_1(x_1)A_2(x_2)A_3(x_3)\rangle=C_{123}\,r_{12}^{X_3-X_1-X_2}r_{13}^{X_2-X_1-X_3}r_{23}^{X_1-X_2-X_3}\,.
\eea
If the fields are chosen and normalized in such a way that $C_{ij}=\delta_{ij}$, the constants $C_{ijk}$, which are invariant under permutations of their indices, coincide with the coefficients $C_{ij}^k$ of the operator product expansion (OPE)
\EQ
A_i(x_1)A_j(x_2)=C_{ij}^k\,r_{12}^{X_k-X_i-X_j}\,A_k(x_2)+\ldots\,.
\label{ope}
\EN

Denoting $X_\sigma$ the scaling dimension of the Potts spin field $\sigma_\alpha(x)$, we see recalling (\ref{p2}-\ref{p3}) that $P_2\propto r^{-2X_\sigma}$, $P_3\propto(r_{12}r_{13}r_{23})^{-X_\sigma}$, and 
\EQ
R(x_1,x_2,x_3)\equiv R_c=\frac{q^3}{(q-1)(q-2)}\,C_{\sigma_\alpha\sigma_\alpha\sigma_\alpha}\,,\hspace{1cm}p=p_c\,,
\label{rc}
\EN
where we have chosen
\EQ
C_{\sigma_\alpha\sigma_\alpha}=\frac{q-1}{q^2}\,,
\label{ssi}
\EN
 in such a way that, in particular, $P_2$ is finite at $q=1$.

In the two-dimensional case, to which we restrict from now on, the critical scaling Potts model with $q\leq 4$ is a conformal field theory \cite{BPZ} with central charge 
\EQ 
c=1-\frac{6}{t(t+1)}\,,
\EN
with $t$ related to $q$ as \cite{DF}
\EQ
\sqrt{q}=2\sin\frac{\pi(t-1)}{2(t+1)}\,.
\label{tq}
\EN
In conformal theories with $c<1$, the formula
\EQ
X_{m,n}=\frac{[(t+1)m-tn]^2-1}{2t(t+1)}
\label{kac}
\EN
determines, for $m$ and $n$ positive integers, the scaling dimensions of scalar ``degenerate'' primary fields having the property that correlation functions containing them satisfy linear differential equations of order $mn$. For some discrete values of $c$, and in particular for any integer $t>2$, there exist ``minimal models'' whose space of fields decomposes into a finite number of subspaces, each one originating from a degenerate primary \cite{BPZ}; the number of primaries with a given scaling dimension is an integer in minimal models \cite{Cardy-mi,CIZ}. In the Potts model, the fields $\sigma_\alpha$ have \cite{Nienhuis,DF} 
\EQ
X_\sigma=X_{(t+1)/2,(t+1)/2}
\label{xsigma}
\EN
and multiplicity $q-1$ ($\sum_\alpha\sigma_\alpha=0)$, allowing minimality only for $q=2$ (Ising model, $t=3$) and $q=3$ ($t=5$); the leading $S_q$-invariant field $\varepsilon$, with $X_\varepsilon=X_{2,1}$, is degenerate for generic values of $q$.

In \cite{DF-ope} the properties of the degenerate fields were exploited to compute the OPE coefficients $C_{ij}^k$ for the ``diagonal'' series of minimal models, in which all the primaries appear with multiplicity one. The Potts model belongs to this class for $q=2$, but even for this case the results for the minimal OPE are not sufficient to determine $R_c$. Indeed, it is clear from (\ref{rc}) that a finite $R_c$ requires $C_{\sigma_\alpha\sigma_\alpha\sigma_\alpha}=0$ at $q=2$, and this is trivially ensured by the spin reversal symmetry of the Ising model. Hence, even the determination of $R_c$ at $q=2$ requires the computation of the OPE coefficient for continuous values of $q$; even forgetting the problem with the multiplicity of the fields, the formulae of \cite{DF-ope} can be evaluated only for the discrete values of $t$ corresponding to minimal models.

Few years ago Al. Zamolodchikov \cite{Alyosha} approached the problem of the derivation of the OPE coefficients of $c<1$ minimal models within a conformal bootstrap method based on the use of correlators of four scalar fields, one of which being the degenerate primary with dimension $X_{1,2}$ or $X_{2,1}$. The other three fields are simply required to appear with multiplicity one; the mathematical treatment does not put any constraint on their scaling dimensions $X_i$, $i=1,2,3$. The method leads to functional equations for the OPE coefficients which are related by analytic continuation to the functional equations arising in Liouville theory ($c\geq 25$) \cite{Teschner}. The solution for $c<1$, however, is not an analytic continuation of the Liouville solution, and reads \cite{Alyosha}
\bea
&&C_{X_1,X_2}^{X_3}=C_{X_1,X_2,X_3}=\label{alyosha}\\
&&\frac{A\,\Upsilon(a_1+a_2-a_3+\beta)\Upsilon(a_2+a_3-a_1+\beta)\Upsilon(a_3+a_1-a_2+\beta)\Upsilon(2\beta-\beta^{-1}+a_1+a_2+a_3)}
{[\Upsilon(2a_1+\beta)\Upsilon(2a_1+2\beta-\beta^{-1})\Upsilon(2a_2+\beta)\Upsilon(2a_2+2\beta-\beta^{-1})\Upsilon(2a_3+\beta)\Upsilon(2a_3+2\beta-\beta^{-1})]^{\frac12}},\nonumber
\eea
with
\bea
&& \beta=\sqrt{t/(t+1)}\,,\\
&& X_i=2a_i(a_i+\beta-\beta^{-1})\,,\label{xa}\\
&& A=\frac{\beta^{\beta^{-2}-\beta^2-1}[\gamma(\beta^2)\gamma(\beta^{-2}-1)]^{1/2}}{\Upsilon(\beta)}\,,\hspace{1.5cm}\gamma(x)\equiv\frac{\Gamma(x)}{\Gamma(1-x)}\,,
\eea
\EQ
\Upsilon(x)=\exp\left\{\int_0^\infty\frac{dt}{t}\left[\left(\frac{Q}{2}-x\right)^2e^{-t}-\frac{\sinh^2\left[\left(\frac{Q}{2}-x\right)\frac{t}{2}\right]}{\sinh\frac{\beta t}{2}\sinh\frac{t}{2\beta}}\right]\right\},\hspace{1cm}Q=\beta+\beta^{-1}.
\label{upsilon}
\EN
The integral in (\ref{upsilon}) is convergent for $0<x<Q$; outside this range $\Upsilon(x)$ can be computed using the relations 
\bea
\Upsilon(x+\beta)&=&\gamma(\beta x)\beta^{1-2\beta x}\Upsilon(x)\,,\\
\Upsilon(x+1/\beta)&=&\gamma(x/\beta)\beta^{2x/\beta-1}\Upsilon(x)\,.
\eea
The function (\ref{alyosha}) can be evaluated for continuous values of $t$ and of the $X_i$'s. Taken literally, it would give the OPE coefficients for a theory with arbitrary $c<1$ and with a continuous spectrum of fields having multiplicity one. It is remarked in \cite{Alyosha} that the consistency of such a theory, in particular from the point of view of modular invariance \cite{Cardy-mi,CIZ}, is an open question. Concerning the values of $c$ corresponding to minimal models, (\ref{xa}) gives the scaling dimensions (\ref{kac}) for $a_i$ equal to
\EQ
a_{m,n}=\frac{(n-1)\beta}{2}-\frac{(m-1)\beta^{-1}}{2}\,.
\EN
Although a general proof is not available, checks case by case show that (\ref{alyosha}) reproduces the OPE coefficients of minimal models obtained in \cite{DF-ope}, at least when these differ from zero. In some cases for which the minimal OPE prescribes vanishing coefficients, (\ref{alyosha}) gives instead finite numbers whose interpretation is considered ``mysterious'' in \cite{Alyosha}. As an example, (\ref{alyosha}) evaluated for $X_1=X_2=X_3=X_{2,2}$ at $t=3$ does not vanish, despite the fact that the spin three-point function is zero in the Ising model.

Recall now that the two-dimensional Potts model contains also disorder fields, dual to the spin fields and with the same scaling dimension $X_\sigma$. They satisfy the OPE
\EQ
\mu_{\alpha\beta}\,\mu_{\beta\gamma}=\delta_{\alpha\gamma}(I+C_\varepsilon\,\varepsilon+\ldots)+(1-\delta_{\alpha\gamma})(C_\mu\,\mu_{\alpha\gamma}+\ldots)\,,
\label{ope-mu}
\EN
where we omit the coordinate dependence for simplicity and the coefficient in front of the identity is fixed to 1 by (\ref{d2}) and (\ref{ssi}). Use of (\ref{d2}-\ref{d3}) at the self-dual point then leads to the result $R_c=C_\mu$. $S_q$-invariance gives to (\ref{ope-mu}) a two-channel structure ($\alpha=\gamma$ or $\alpha\neq\gamma$) equivalent to that produced by two fields $\mu$ and $\bar{\mu}$ satisfying
\bea
\mu\,\bar{\mu}&=&I+C_\varepsilon\,\varepsilon+\ldots\,,\\
\mu\,\mu+\bar{\mu}\,\bar{\mu}&=&C_\mu(\mu+\bar{\mu})+\ldots\,.
\eea
The field $\phi\equiv(\mu+\bar{\mu})/\sqrt{2}$ then satisfies
\EQ
\phi\,\phi=I+C_\varepsilon\,\varepsilon+\frac{C_\mu}{\sqrt{2}}\,\phi+\ldots\,,
\label{ope-phi}
\EN
namely a ``neutral'' OPE for fields with multiplicity one, as the one assumed in the derivation of (\ref{alyosha}). Notice also that, due to the neutrality of the fields in (\ref{ope-phi}), $C_\mu$ has no reason to vanish for any value of $q$; in particular, this is not in conflict with the symmetries of the Ising model, because the absence for $q=2$ of the vertex with three disorder fields in (\ref{ope-mu}) is enforced by the factor $1-\delta_{\alpha\gamma}$. 

These observations may suggest the following interpretation for the function (\ref{alyosha}): it encodes the ``dynamical'' information about the OPE of conformal field theory for $c<1$, and knows nothing about internal symmetries, which, on the other hand, are not uniquely determined by the value of $c$; symmetry considerations have to be developed separately and produce a dressing of (\ref{alyosha}) by factors which may vanish, suppressing the vertices not compatible with the given symmetry. Letting aside the general validity of this interpretation, the OPE's (\ref{ope-mu}) and (\ref{ope-phi}) make it plausible for the case we are discussing, and lead us to take 
\EQ
R_c=\sqrt{2}\,C_{X_\sigma,X_\sigma,X_\sigma}\,,
\label{result}
\EN
with $X_\sigma$ given in (\ref{xsigma}). The results for the integer values of 
$q$ are given in Table~\ref{table1}; the value for $q=4$ is obtained in the limit $t\to\infty$. In \cite{SZK} the constant value $\approx 1.022$ was obtained numerically for critical percolation on a cylinder and (\ref{r}) evaluated for $x_1$ on one edge, $x_3$ on the other edge and $x_2$ far away from both edges. Since in such a configuration the cylinder, seen from $x_2$, looks infinitely long and becomes conformally equivalent to the plane, we see in this result a confirmation of our analytic value for $R_c$ at $q=1$.

\begin{table}[htbp]
\begin{center}
\begin{tabular}{|c|c|c|c|c|}
\hline
$q$ & 1 & 2 & 3 & 4 \\
\hline
$R_c$ & $1.0220..$ & $1.0524..$ & $1.0923..$ & $1.1892..$ \\
\hline
$\tilde{R}_c$ & $-$ & $1.3767..$ & $1.3107..$ & $1.1892..$ \\
\hline
$\Gamma_{KK}^K$ & $1.0450..$ & $1.1547..$ & $1.3160..$ & $1.8612..$ \\
\hline
\end{tabular}
\caption{$R_c$ and $\Gamma_{KK}^K$ are the values of the quantities (\ref{rc}) and (\ref{vertex}) for KF clusters; $q=1$ corresponds to random percolation. $\tilde{R}_c$ is obtained from (\ref{result}) with $X_\sigma$ replaced by $X_{\tilde{\sigma}}$, the scaling dimension associated to spin clusters.} \label{table1}
\end{center}
\end{table}

The result (\ref{result}) refers to KF clusters. Concerning the ordinary spin clusters, i.e. those obtained connecting nearest neighbors with the same value of the spin, they are also critical at $J_c$ in two dimensions \cite{CP}, with connectivities related to the correlation functions of the field with scaling dimension $X_{\tilde{\sigma}}=X_{t/2,t/2}$ \cite{Vanderzande,DJS}. For the Ising case, the best understood field theoretically \cite{isingperc,DV}, the ratio (\ref{r}) is expected to be given by (\ref{result}) with $X_\sigma$ replaced by $X_{\tilde{\sigma}}$. We give this value $\tilde{R}_c$ in Table~\ref{table1}; those obtained in the same way at $q=3,4$ are also quoted, but these cases are less clear. 

Comparison between (\ref{rc}) and (\ref{result}) determines $C_{\sigma_\alpha\sigma_\alpha\sigma_\alpha}$ and its relation with $C_\mu$. The relations $C_{\sigma_\alpha\sigma_\beta}=(q\delta_{\alpha\beta}-1)/q^2$ and $C_{\sigma_1\sigma_1\sigma_1}=(1-q)C_{\sigma_1\sigma_1\sigma_2}=\frac12(q-1)(q-2)C_{\sigma_1\sigma_2\sigma_3}$ are a consequence of the symmetry. Eqs.~(\ref{d2}) and (\ref{ope-phi}) lead to $C_\varepsilon=-q^2(q-1)^{-1}C_{\sigma_\alpha\sigma_\alpha}^\varepsilon=C_{X_\sigma,X_\sigma,X_\varepsilon}$, where we also took into account that $\varepsilon$ changes sign under duality\footnote{For example (\ref{alyosha}) gives $C_\varepsilon|_{q=2}=-1/2$.}. The OPE  between  $\sigma_\alpha$ and $\mu_{\beta\gamma}$ produces parafermionic fields. An analysis similar to that of \cite{lambda} suggests that they have spin $X_{1,3}/2$; this is certainly the case for $q=2,3$ \cite{ZF}.

\vspace{.3cm}
{\bf 3.} Away from criticality the Potts field theory is solved exactly in the framework of the factorized $S$-matrix \cite{CZ}, from which large distance expansions can be obtained for the correlation functions \cite{DC,DVC}. The $S_q$ symmetry is more transparently implemented working in the ordered phase ($J>J_c$), where the elementary excitations (kinks) are interpolated by the disorder fields; the results for the disordered phase are obtained by duality. If $x_1,x_2,x_3$ are the vertices of a triangle whose internal angles are all smaller than $2\pi/3$, the asymptotic behavior of the correlator (\ref{d3}) when all the distances between the vertices become large reads \cite{CZ} (\cite{CDGJM} for a derivation)

\begin{figure}
\begin{center}
\includegraphics[width=4cm]{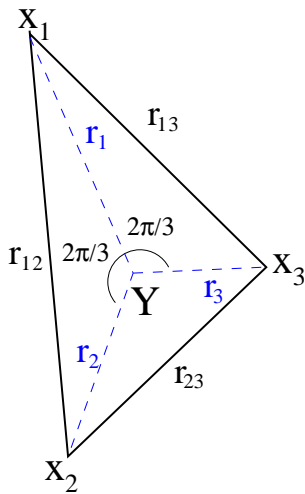}
\caption{Triangle identified by the points $x_1, x_2, x_3$. The lines joining the vertices of the triangle to the Fermat (or Steiner) point $Y$ form $2\pi/3$ angles.}
\label{triangle}
\end{center} 
\end{figure}

\EQ
\label{mu3}
\langle\mu_{\alpha\beta}(x_1)\mu_{\beta\gamma}(x_2)\mu_{\gamma\alpha}(x_3)\rangle_{J>J_c}=\frac{F^3_\mu}{\pi}\,\Gamma_{KK}^K\,K_0(r_{Y}/\xi)+O(\text{e}^{-\rho/\xi})\,.
\EN
Here $F_\mu$ is the one-kink form factor of the disorder field, $K_0(x)=\int_0^{\infty}\text{d}y\,\text{e}^{-x\cosh y}$ is a Bessell function, and $r_Y\equiv r_1+r_2+ r_3$ is the sum of the distances of the vertices of the triangle from the Fermat (or Steiner) point $Y$, which has the property of minimizing such a sum (Fig.~\ref{triangle}); $r_{Y}<\rho\equiv\text{min}\{r_{12}+r_{13},r_{12}+r_{23},r_{13}+r_{23}\}$. $\xi$ is the connectivity length (inverse of the kink mass), also determined by the large distance decay of the two-point function 
\EQ
\langle\mu_{\alpha\beta}(x_1)\mu_{\beta\alpha}(x_2)\rangle_{J>J_c}=\frac{F_{\mu}^2}{\pi}\,K_0(r_{12}/\xi)+O(\text{e}^{-2r_{12}/\xi})\,,
\label{mu2}
\EN
and $\Gamma_{KK}^K$ is the three-kink vertex given by \cite{CZ}
\bea
\label{vertex}
& &\Gamma_{KK}^K=\sqrt{\frac{1}{\lambda}\sin\frac{2\pi\lambda}{3}\,g(\lambda)}\,,\\\label{gfunction}
& & g(\lambda)=\text{exp}\left[\int_{0}^{\infty}\text{d}t
  \sinh\left(\frac{t}{3}\right)\frac{\sinh\left[\frac{t}{2}\left(1-\frac{1}{\lambda}\right)\right]-
\sinh\left[\frac{t}{2}\left(\frac{1}{\lambda}-\frac{5}{3}\right)\right]}{t\sinh\frac{t}{2\lambda}\cosh\frac{t}{2}} \right]\,,
\eea
with $q=2\sin\frac{\pi\lambda}{3}$, $\lambda\in(0,3/2)$. It follows from (\ref{d2}-\ref{d3}) that (\ref{mu3}-\ref{mu2}) are precisely the asymptotics for large separations of the connectivties (\ref{p2}-\ref{p3}) for KF clusters below $p_c$. It follows
\EQ
R(x_1,x_2,x_3)\simeq\sqrt{\pi}\,\Gamma_{KK}^K\,\frac{K_0(r_{Y}/\xi)}{\sqrt{K_0(r_{12}/\xi)K_0(r_{13}/\xi)K_0(r_{23}/\xi)}}\,,\hspace{1cm}{r_{ij}}\gg{\xi},\hspace{.5cm}p\to p_c^-\,.
\label{asymp}
\EN
The dynamical information is entirely contained in the three-kink vertex, whose values for $q$ integer are given\footnote{The value $\Gamma_{KK}^K|_{q=1}$ quoted in \cite{CZ} is affected by a typo in the formula corresponding to our (\ref{gfunction}).} in Table~\ref{table1}. In the opposite limit, in which all the distances $r_{ij}$ are much smaller than $\xi$ (always remaining much larger than the lattice spacing), the function $R$ tends to its constant critical value $R_c$.

Finally, consider the ordinary spin clusters for the Ising model. It was argued in \cite{isingperc} that in this case the scaling limit for $p\to p_c^-$ (in zero magnetic field) corresponds to a renormalization group trajectory with infinite connectivity length ending into a random percolation fixed point at large distances. Then one expects (\ref{r}) to interpolate from $\tilde{R}_c|_{q=2}$, when all the distances between the points are small, to $R_c|_{q=1}$, when all of them are large. Although universal, this crossover could be easier to observe on the triangular lattice, which has $p_c=1/2$ for site percolation and is expected to minimize the corrections to scaling \cite{isingperc}. If instead we work at the Ising critical temperature and add a magnetic field $H$, the connectivity length is finite and the large separation behavior of (\ref{r}) for clusters of positive spins and $H\to 0^-$ has again the form (\ref{asymp}), with $\Gamma_{KK}^K=5.7675..$ given by (\ref{vertex}) evaluated at $\lambda=5/2$ \cite{DV}. Since the integral in (\ref{gfunction}) diverges for $\lambda\geq 3/2$, one uses the analytic continuation
\EQ
g(\lambda)=\frac{\Gamma\left(1-\frac{2\lambda}{3}\right)\Gamma(1+\lambda)}{\Gamma\left(1+\frac{\lambda}{3}\right)}\,
\text{exp}\left[\int_{0}^{\infty}\text{d}t\sinh\left(\frac{t}{3}\right)\frac{\sinh\left[\frac{t}{2}\left(1-\frac{1}{\lambda}\right)\right]-\text{e}^{-t}\sinh\left[\frac{t}{2}\left(\frac{1}{3}+\frac{1}{\lambda}\right)\right]}{
t\sinh\frac{t}{2\lambda}\cosh\frac{t}{2}} 
\right]\nonumber
\EN
for $3/2\leq\lambda<3$.


\vspace{1cm} \textbf{Acknowledgments.} We thank J. Cardy, P. Kleban, J. Simmons and R. Ziff for discussions. GD thanks NORDITA (Stockholm) for hospitality during the final stages of this work. Work supported in part by ESF Grant INSTANS and by MIUR Grant 2007JHLPEZ.

\end{document}